\documentstyle[a4wide,epsf,11pt,titlepage]{article}

\newcounter{nref}
\newcommand{\bbib}{%
  \renewcommand{\refname}{\large\bf References}%
  \begin{thebibliography}{99}%
  \setcounter{enumiv}{\thenref}}
\newcommand{\ebib}{%
  \end{thebibliography}%
  \setcounter{nref}{\arabic{enumiv}}}
\newcommand{\head}[3]{%
  \setcounter{nref}{0}%
  \thispagestyle{empty}%
  \section*{\LARGE\bf #1}%
  \stepcounter{section}%
  \addcontentsline{toc}{section}{#1}%
  \large\itshape%
  #2\\\vspace{0.1pt}\\%
  #3%
  \normalsize\upshape%
  \bigskip}

\def\simge{\, {}^>_{\sim }\,}
\def\simle{\, {}^<_{\sim }\,}

\begin{document}


\head{Non-Standard Big Bang Nucleosynthesis Scenarios}
     {K.\ Jedamzik}
     {Max-Planck-Institut f\"ur Astrophysik, Karl-Schwarzschild-Str.1,
     D--85748 Garching\\  }

\subsection*{Abstract}

A brief overview of non-standard big bang nucleosynthesis (BBN) scenarios
is presented. 
Trends and results of the light-element nucleosynthesis in BBN
scenarios with small-scale or large-scale inhomogeneity, the
presence of antimatter domains, stable or unstable massive
neutrinos, neutrino oscillations, neutrino degeneracy, or massive decaying
particles are summarized.

\subsection*{Introduction}

Light-element nucleosynthesis during the BBN epoch below
cosmic temperatures $T\approx 1$MeV
contributes significantly to $^4$He, $^3$He and $^7$Li
abundances and likely, to all of the $^2$H abundance observed 
throughout the universe.
BBN is a freeze-out process from nuclear statistical equilibrium such
that light-element abundance yields are sensitively dependent on
the cosmic conditions during the BBN era as well as the properties
of neutrinos governing the freeze-out process from weak equilibrium.
Calculations of abundance yields in a standard BBN scenario are performed
under the assumptions of a universe homogeneous
in the baryon-to-photon ratio, with massless 
neutrinos and vanishing neutrino chemical potentials, and in the
absence of massive decaying particles or other degrees of freedom.
For reviews on the physics of standard BBN see \cite{kj:skm93}.
The purpose of this short article is to summarize results and trends
in theoretical calculations of non-standard BBN scenarios where one 
of the above assumptions in standard BBN is relaxed.
This summary is in no way intended to be complete neither in 
discussing all possible modifications to a standard BBN scenario
nor in referencing the thousands of articles on non-standard BBN.
I wish to apologize for including only certain key references due
to the limited scope of these proceedings and for presenting my
personal view of the field of non-standard BBN.
For two excellent reviews on non-standard BBN the reader is
referred to \cite{kj:mm93}.
The determination of observationally inferred
primordial abundance constraints, which represents possibly the most
important branch of the field of big bang nucleosynthesis at present,
will not be discussed here. In what follows, abundance yields in non-standard
BBN will either be given in relation to abundance yields in standard BBN
or in absolute values and shall be understood as indicative of approximate trends.

\subsection*{Non-Standard BBN}

In the following a list of non-standard BBN scenarios, their respective modifications
to standard BBN (hereafter; SBBN), 
as well as trends and results in these scenarios are given.

\subsubsection*{Inhomogeneity}

The baryon-to-photon ratio $\eta$ is the one free parameter
in SBBN. Any inhomogeneity in this quantity results
in modified nucleosynthesis yields which depend on the typical 
amplitude and spatial seperation scale of inhomogeneities.
Substantial changes in the abundance yields only result
when $\delta\eta /\eta \simge 1$.

\vskip 0.1in
\noindent
{\bf a) Inhomogeneity in the Baryon-to-Photon Ratio on Small Mass
Scales:}
\vskip 0.05in

\noindent
Fluctuations in $\eta$  which may arise on sub-horizon scales at
earlier cosmic epochs as, for example, possibly
during a first-order QCD transition or electroweak transition,
result in a highly nonstandard BBN scenario.
The nucleosynthesis in an environment with $\eta$-fluctuations is characterized
by coupled nuclear reactions and hydrodymamic processes, such as
baryon diffusion and late-time expansion of high-density regions \cite{kj:ahs87}.
Fluctuations in $\eta$ persist down to the onset of BBN provided
the mass of an individual high-density region exceeds $10^{-21}M_{\odot}$.
One of the main features of such scenarios is the differential diffusion
of neutrons and protons leading to the existence of neutron-
and proton-rich environments. The trend in inhomogeneous BBN 
is the overabundant production of $^4$He when compared
to SBBN at the same average $\eta$, nevertheless, there exists
parameter space where less $^4$He than in SBBN is synthesized.
Scenarios which don't overproduce $^4$He typically have high 
($^2$H/H) $\sim 1-2\times 10^{-4}$
and high ($^7$Li/H) $\sim 10^{-9} - 10^{-8}$. Such BBN may agree
with observational abundance constraints for 
fractional contributions of the baryon density to the
critical density $\Omega_b$ about 2-3
times larger than in SBBN, but only in the seemingly unlikely advent
of efficient $^7$Li depletion in population II stars.

\vskip 0.1in
\noindent
{\bf b) Inhomogeneity in the Baryon-to-Photon Ratio on 
Large Mass Scales:}
\vskip 0.05in

\noindent
When the baryonic mass within a typical fluctuation
exceeds ($M \simge 10^{-12}M_{\odot}$), baryon diffusion and
hydrodymanic processes during the BBN era are of
no significance such that BBN abundance yields may
be given by an average over the SBBN abundance yields of seperate regions
at different $\eta$. For non-linear fluctuations exceeding the
post-recombination
Jeans mass $M \simge 10^5M_{\odot}$, which may exist in
primordial isocurvature baryon (PIB) models for structure formation, early
collapse
of high-density ($\eta$) regions is
anticipated \cite{kj:sm86}. The nucleosynthesis yields of collapsing regions
may be excluded from the primordial abundance determination
if either dark objects form or significant early star formation
in such high-density regions occurs.  
If only low-density regions contribute to the observable
primordial abundances, characteristic
average abundance yields for scenarios designed to possibly
agree with observationally inferred primordial abundances are:
($^2$H/H) $\sim 1-3\times 10^{-4}$, ($^{7}$Li/H) $\sim 5\times 10^{-10} - 2\times
10^{-9}$,
and $^4$He mass fraction $Y_p\approx 0.22-0.25$ at a 
total $\Omega_b \simle 0.2$ (i.e. including possible dark obje
cts), 
larger than inferred from a SBBN scenario \cite{kj:jf95,kj:cos95}.
One feature of such models is the prediction of
fairly large intrinsic spatial variations in the primordial abundances,
which may be observationally tested by ($^2$H/H) determinations in Lyman-limit
systems \cite{kj:jf95}.
These models may only agree with observationally inferred abundance limits
when there
are no
fluctuations below $M \simle 10^5M_{\odot}$ and collapse efficiencies
of high-density regions are large.

\vskip 0.1in
\noindent
{\bf c) Matter/Antimatter Domains:}
\vskip 0.05in

\noindent
A distribution of small-scale matter/antimatter domains in
baryon-asymmetric universes (i.e. where net $\eta \neq 0$) may result from electroweak
baryogenesis scenarios. If the baryon (antibaryon) mass in individual domains
is $\simge 10^{-21}M_{\odot}$ the BBN process in such scenarios is characterized by
differential diffusion of neutrons (antineutrons) and protons (antiprotons) 
which causes a preferential annihilation of antimatter on neutrons
\cite{kj:St76}. When annihilation of antimatter occurs before
significant $^4$He synthesis ($T\simge 80$keV) 
but after weak freeze-out ($T\simle 1$MeV)
a modest to substantial reduction of $Y_p$ results.
When annihilations occur mainly after $^4$He synthesis the dominant effect is
significant production of $^3$He and $^2$H
\cite{kj:betal88}, with $^3$He/$^2$H ratios likely to be
in conflict with observational constraints.

\subsubsection*{Non-standard Neutrino Properties}

The BBN process may be approximated as the incorporation of all available
neutrons into $^4$He nuclei at a temperature $T\approx 80$keV.
The neutron abundance at this temperature, and hence the final 
$^4$He mass fraction $Y_p$, may
be increased with respect to a SBBN scenario due to an increased
expansion rate of the universe during the BBN era. An increased
expansion rate raises the 
neutron-to-proton ratio (hereafter; n/p) at weak freeze-out 
and reduces the time for neutron decay to decrase the 
n/p ratio
between weak freeze-out and significant $^4$He synthesis.
We will refer to this effect as the \lq\lq expansion rate effect\rq\rq .
In addition, the n/p ratio at $T\approx 80$keV may be either decreased or increased
by introducing additional
electron- and/or anti-electron neutrinos into the plasma.
In SBBN it is assumed that the left-handed neutrino and right-handed
antineutrino seas of three massless, stable neutrino flavors $\nu_e$, $\nu_{\mu}$, 
and $\nu_{\tau}$
are populated and that neutrino chemical potentials do vanish.
Modifications of these assumptions usually result in either the expansion
rate effect
or additional (anti) electron neutrinos, or both. The principal effect of
such modifications is to change the $^4$He abundance, and to a less 
observationally significant
degree the abundances of other light-element isotopes.

\vskip 0.1in
\noindent
{\bf a) Massive, long-lived $\tau$-neutrinos:}
\vskip 0.05in

\noindent
Neutrinos are considered massless and long-lived in the context of BBN
for neutrino masses $m_{\nu}\simle 100$keV and lifetimes $\tau_{\nu}\simge 
10^3$s 
(see b), however, possible photodisintegration). A massive, long-lived
$\tau$-neutrino leads to the expansion rate effect since
the contribution to the total energy density from the rest mass
of $\tau$-neutrinos continually increases as the universe expands between
weak freeze-out and $^4$He synthesis, possibly even resulting in matter
domination during the BBN era. BBN with massive, long-lived $\tau$ neutrinos
and for experimentally allowed $\nu_{\tau}$-masses therefore results in
increased $^4$He and useful limits on the allowed
mass of a long-lived $\tau$-neutrino have been derived
\cite{kj:ketal91}.

\vskip 0.1in
\noindent
{\bf b) Massive, unstable $\tau$-neutrinos:}
\vskip 0.05in

\noindent
The effects of decaying $\tau$ neutrinos on the light-element nucleosythesis
\cite{kj:ks82}
sensitively depend on the decay products. One distinguishes between
(i) decay into sterile particles, in particular, particles which interact
neither weakly with nucleons interchanging neutrons and protons
nor electromagnetically with the ambient plasma
(e.g. $\nu_{\tau}\mapsto \nu_{\mu} + \phi$, where $\phi$ is a weakly interacting
scalar), 
(ii) decay into sterile and electromagnetically interacting particles
(e.g. $\nu_{\tau}\mapsto \nu_{\mu} + \gamma$), 
and (iii) decay into (anti) electron neutrinos and sterile
particles (e.g. $\nu_{\tau}\mapsto \nu_{e} + \phi$ )
\cite{kj:dgt94,kj:ketal94}. 
For decay channel (i) and $\tau_{\nu} \simge 1$s increased $^4$He mass fraction
results due
to the expansion rate effect which, nevertheless, is weaker than for
long-lived $\tau$-neutrinos since the energy 
of the (massless) decay products redshifts with
the expansion of the universe. 
In contrast, for life times 
$\tau_{\nu_{\tau}} \simle 1$s and $m_{\nu_{\tau}}\simge 10$MeV
it is possible to reduce the $Y_p$
since effectively the distributions of only two neutrino flavors are populated
\cite{kj:kks97}.
Decay channel (ii) would have interesting
effects
on BBN but is excluded by observations of supernova 1987A for $\tau_{\nu_{\tau}}
\simle 10^4$s.
For decay via channel (iii) additional (anti) electron neutrinos are injected
into the plasma
and their effect depends strongly on the time of injection \cite{kj:ts88}.
When injected early ($\tau_{\nu_{\tau}}\sim 1$s), the net result
is a reduction of $Y_p$ \cite{kj:h97} 
since weak freeze-out occurs at lower temperatures.
In contrast, when injected late ($\tau_{\nu} \sim 10^2 - 10^3$s) the resulting 
non-thermal
electron neutrinos affect a conversion of protons into neutrons,
yielding higher $Y_p$ and/or higher $^2$H depending on injection time. 
It had been suggested that a scenario with late-decaying $\nu_{\tau}$ via channel
(iii) may result in the relaxation of BBN bounds on $\Omega_b$ by a factor
up to ten \cite{kj:dgt94}, nevertheless, 
this possibility seems to be ruled out now by the
current upper laboratory limit on the $\nu_{\tau}$-mass.
For life times $\tau_{\nu_{\tau}}\simge
10^3$s, modifications
of the light-element abundances after the BBN era by photodisintegration of
nuclei may result (cf. radiative decays).

\vskip 0.1in
\noindent
{\bf c) Neutrino oscillations:}
\vskip 0.05in

\noindent
Neutrino oscillations may occur when at least one neutrino species
has non-vanishing mass and the weak neutrino interaction eigenstates
are not mass eigenstates of the Hamiltonian. One distinguishes between
(i) flavor-changing neutrino oscillations 
(e.g. $\nu_{e}\longleftrightarrow \nu_{\mu}$) and (ii) active-sterile
neutrino oscillations (e.g. $\nu_{e}\longleftrightarrow \nu_{s}$). 
Here $\nu_s$ may be either the right-handed
component of a $\nu_e$ ($\nu_{\mu}$, $\nu_{\tau}$) Dirac neutrino, 
or a fourth family of neutrinos
beyond the standard model of electroweak interactions. In the absence of
sterile neutrinos and neutrino degeneracy, neutrino oscillations
have negligible effect on BBN due to the almost equal number densities of
neutrino
flavors. When sterile neutrinos exist, neutrino oscillations may result into
the population of the sterile neutrino distribution, increasing the
energy density, and leading to the expansion rate effect. The increased $Y_p$
has been used to infer limits on the neutrino squared-mass difference --
mixing angle
plane \cite{kj:bd91}. 
In the presence of large initial lepton number asymmetries 
(see neutrino degeneracy) and with sterile neutrinos, it may be possible to reduce $Y_p$
somewhat
independently from the detailled initial conditions, 
through the dynamic generation
of electron (as well as, $\mu$ and $\tau$) neutrino chemical potentials \cite{kj:fv}.

\subsubsection*{Neutrino Degeneracy}

It is possible that the universe has net lepton number.
Positive net cosmic lepton number manifests itself at low temperatures
through an excess of neutrinos over antineutrinos. If net lepton number
in either of the three families in the standard model is about ten orders
of magnitude larger than the net cosmic baryon number BBN abundance yields
are notabely affected. Asymmetries between the $\nu_{\mu}$ ($\nu_{\tau}$) and 
$\bar{\nu}_{\mu}$ ($\bar{\nu}_{\tau}$)
number densities result in the expansion rate effect only, whereas
asymmetries between the $\nu_e$ and $\bar{\nu}_e$ number densities induce a change
in the weak freeze-out (n/p) ratio as well.
Since the expansion rate effect leads to increased $^4$He production
$\nu_{\mu}$ ($\nu_{\tau}$) degeneracy may be constrained.
However, one may find combinations
of $\nu_{\mu}$, $\nu_{\tau}$, {\it and} $\nu_e$
chemical potentials which
are consistent with observational abundance constraints for $\Omega_b$ much
larger than that inferred from SBBN \cite{kj:by77}. 
Nevertheless, such solutions not only
require large chemical potentials but also an asymmetry between the
individual chemical potentials of $\nu_e$ and $\nu_{\mu}$ ($\nu_{\tau}$). 
Asymmetries
between the different flavor degeneracies may be erased in the presence of
neutrino oscillations.

\subsubsection*{Massive Decaying Particles}

The out-of-equilibrium decay or annihilation of long-lived particles
($\tau \simge 0.1$s), such
as light supersymmetric particles, as well as non-thermal particle production
by, for example, evaporating primordial black holes or collapsing cosmic
string loops, during, or after, the BBN era may significantly alter the
BBN nucleosynthesis yields \cite{kj:detal78}.
The decays may be classified according to if they are (i) radiative or
(ii) hadronic, in particular, whether the electromagnetic
or strong interactions of the injected non-thermal particles are most
relevant.

\vskip 0.1in
\noindent
{\bf a) Radiative decays:}
\vskip 0.05in

\noindent
Electromagnetically interacting particles (i.e.$\gamma$, $e^{\pm}$) thermalize
quickly in the ambient photon-pair plasma at high temperatures, such
that they usually have little effect on BBN other than some heating
of the plasma. In contrast, if radiative decays occur
at lower temperatures after a conventional BBN
epoch, they result in a rapidly developing $\gamma$-$e^{\pm}$ cascade which only
subsides when individual $\gamma$-rays of the cascade do not have enough energy 
to further
pair-produce $e^{\pm}$ on the ambient cosmic background radiation. The net result
of this cascade is a less rapidly developing $\gamma$-ray background
whose properties only depend on the total amount of energy released 
in electromagnetically interacting particles and the epoch of decay.
At temperatures $T\simle 5$keV the most energetic $\gamma$-rays in this background may 
photodissociate
deuterium, at temperatures $T\simle 0.5$keV the photodisintegration of 
$^4$He becomes possible.
Radiative decays at $T\approx 5$keV may result in a BBN scenario with low $^2$H
{\it and} low $^4$He if an ordinary SBBN scenario at low $\eta$ 
is followed by an epoch of
deuterium photodisintegration \cite{kj:hkm96}. 
Radiative decays at lower temperatures may
produce signficant amounts of $^2$H and $^3$He. Nevertheless, this process is unlikely
to be the sole producer of $^2$H due to resulting observationally disfavored
$^3$He/$^2$H ratio. The possible overproduction of $^3$He and $^2$H by the
photodisintegration
of $^4$He has been used to place meaningful limits on the amount of non-thermal,
electromagnetically interacting energy
released into the cosmic background. These limits may actually be
more stringent than comparable limits from the distortion of the cosmic
microwave background radiation \cite{kj:setal96}.

\vskip 0.1in
\noindent
{\bf b) Hadronic decays:}
\vskip 0.05in

\noindent
The injection of hadrons into the plasma, such as $\pi^{\pm}$'s, $\pi^0$'s 
and nucleons, may
affect light-element nucleosythesis during, or after, the BBN era and by
the destruction {\it and} production of nuclei. In general, possible scenarios and
reactions are numerous. If charged hadrons are
produced with high energies below cosmic temperature $T\simle $keV
they may cause a cascade leading
to the possibility of photodisintegration of nuclei (see radiative decays).
Through charge exchange reactions the release of about ten pions per nucleon
at $T\approx 1$MeV results in a significant perturbation of weak freeze-out and
increased $Y_p$ \cite{kj:rs88}. 
A fraction of only $\sim 10^{-3}$ antinucleons per nucleon may cause
overproduction of $^3$He and $^2$H through antinucleon -- $^4$He annihilations.
It is also conceivable that the particle decaying carries baryon number such
that the cosmic baryon number is created {\it during} the BBN era.
A well-studied BBN scenario is the injection of high-energy ($\sim 1$GeV) nucleons
created in hadronic jets which are produced by the decay of the parent particle
\cite{kj:detal88}.
If released after a conventional BBN era these
high-energy nucleons may spall pre-existing $^4$He, thereby producing
high-energy $^2$H, $^3$H, and $^3$He as well as neutrons. Such
energetic light nuclei may initiate
an epoch of non-thermal nucleosythesis.
It was found that the nucleosynthesis yields from a BBN scenario
with hadro-destruction/production and photodisintegration may result in
abundance yields independent of $\eta$ for a range in total energy
injection, and half-life of the decaying particles. Nevertheless, such
scenarios seem to produce $^6$Li in conflict with observational constraints.

\subsubsection*{Other Modifications to BBN}

There are many other variants to a standard big bang nucleosythesis scenario
which have not been mentioned here.
These include anisotropic expansion, variations of fundamental constants,
theories other than general relativity, magnetic fields during BBN,
superconducting cosmic strings during BBN, among others.
The influence of many, but not all, of those scenarions on BBN is due to
the expansion
rate effect. Studies
of such variants is of importance mainly due to the constraints they allow one to 
derive on the evolution of the early universe.


\end{document}